\documentclass[conference, 10pt]{IEEEtran}

\ifCLASSINFOpdf
  \usepackage[pdftex]{graphicx}
   \graphicspath{{..1/pdf/}{../jpeg/}}
   \DeclareGraphicsExtensions{.pdf,.jpeg,.png}
\else
   \usepackage[dvips]{graphicx}
   \graphicspath{{../eps/}}
   \DeclareGraphicsExtensions{.eps}
\fi

\usepackage{cite}
\usepackage{comment} 
\usepackage[cmex10]{amsmath}
\usepackage{amsfonts, amsmath, amsthm, amstext, amssymb, mathrsfs, graphicx, color, url}
\usepackage{array}
\usepackage{mdwmath}
\usepackage{mdwtab}
\usepackage{algpseudocode}
\usepackage{algorithm}
\usepackage{url}
\usepackage{setspace}
\usepackage{footnote}
\usepackage[skip=0pt]{caption}
\usepackage{eqparbox}
\usepackage[tight,footnotesize]{subfigure}
\usepackage{fixltx2e}

\newcommand{\bA}{  \pmb{A}  }

\newcommand{\bD}{  \pmb{D}  }

\newcommand{\bI}{  \pmb{I}  }
\newcommand{\bF}{  \pmb{F}  }
\newcommand{\bG}{  \pmb{G}  }
\newcommand{\bH}{  \pmb{H}  }
\newcommand{\bQ}{  \pmb{Q}  }

\newcommand{\bT}{  \pmb{T}  }
\newcommand{\bX}{  \pmb{X}  }

\newcommand{\bU}{  \pmb{U}  }

\newcommand{\bW}{  \pmb{W}  }
\newcommand{\bPhi}{  \pmb{\Phi}  }

\newcommand{\bGam}{  \pmb{\Gamma}  }
\newcommand{\bLam}{  \pmb{\Lambda}  }

\newcommand{\bSig}{  \pmb{\Sigma}  }
\newcommand{\bx}{  \pmb{x}  }
\newcommand{\by}{  \pmb{y}  }

\newcommand{\bn}{  \pmb{n}  }
\newcommand{\bz}{  \pmb{z}  }
\newcommand{\bp}{  \pmb{p}  }
\newcommand{\bq}{  \pmb{q}  }

\newcommand{\bs}{  \pmb{s}  }
\newcommand{\be}{  \pmb{e}  }
\newcommand{\ba}{  \pmb{a}  }

\newcommand{\bff}{  \pmb{f}  }
\newcommand{\br}{  \pmb{r}  }
\newcommand{\bo}{  \pmb{0}  }

\newcommand{\tbGam}{ \tilde{ \pmb{\Gamma} } }
\newcommand{\tbPhi}{ \tilde{ \pmb{\Phi} } }
\newcommand{\tbTh}{ \tilde{ \pmb{\Theta} } }
\newcommand{\tbgam}{ \tilde{ \pmb{\gamma} } }

\newcommand{\tbf}{ \tilde{ \pmb{f} } }

\newcommand{\tbs}{ \tilde{ \pmb{s} } }
\newcommand{\tbw}{ \tilde{ \pmb{w} } }

\newcommand{\tbF}{ \tilde{ \pmb{F} } }

\newcommand{\tbH}{ \tilde{ \pmb{H} } }

\newcommand{\st}{ \textup{s. t.} }
\newcommand{\argmin}{ \textup{argmin} }

\newcommand{\calS}{ \mathcal{S} }
\newcommand{\bbC}{ \mathbb{C} }

\newtheorem{lemma}{Lemma}

\newtheorem{proposition}{Proposition}

\newcommand{\ww}[2]{  \pmb{w}^{(#1)}_{#2}  }

\begin{document}

\title{Subspace Estimation and Decomposition for Hybrid Analog-Digital Millimetre-Wave MIMO systems}

\author{\IEEEauthorblockN{Hadi Ghauch\IEEEauthorrefmark{1},
Mats Bengtsson\IEEEauthorrefmark{1}, 
Taejoon Kim\IEEEauthorrefmark{2},
Mikael Skoglund\IEEEauthorrefmark{1}     }

\IEEEauthorblockA{\IEEEauthorrefmark{1}School of Electrical Engineering and the ACCESS Linnaeus Center, Royal Institute of Technology (KTH) }

\IEEEauthorblockA{\IEEEauthorrefmark{2} Department of Electronic Engineering, City University of Hong Kong }    
}

\maketitle

\begin{abstract}
In this work, we address the problem of channel estimation and precoding / combining for the so-called hybrid millimeter wave (mmWave) MIMO architecture. Our proposed channel estimation scheme exploits channel reciprocity in TDD MIMO systems, by using echoing, thereby allowing us to implement Krylov subspace methods in a fully distributed way. The latter results in estimating the right (resp. left) singular subspace of the channel at the transmitter (resp. receiver). Moreover, we also tackle the problem of subspace decomposition whereby the estimated  right (resp. left) singular subspaces are approximated by a cascade of analog and digital precoder (resp. combiner), using an iterative method. Finally we compare our scheme with an equivalent fully digital case and conclude that a relatively similar performance can be achieved, however, with a drastically reduced number of RF chains - $4 \sim 8$ times less (i.e., massive savings in cost and power consumption). 
\end{abstract}


\IEEEpeerreviewmaketitle

\section{Introduction}
Communication in the millimeter wave (mmWave) band is one of the strongest candidates to answer the fundamental challenge of the exponentially increasing demand for data rates, in cellular networks. It has the distinct advantage of exploiting the large unused spectrum bands, thereby \emph{offering up to $200$ times more spectrum} than conventional cellular systems. Furthermore, the resulting antenna size/spacing at such frequencies is extremely small, thus implying that \emph{ a large number of such antennas} can be assumed at both the base station and the user (e.g. tens to hundreds). The so-called \emph{hybrid architecture}, first reported in ~\cite{Sayeed_CAP_10, Venkateswaran_analogBF_10}, and later studied in \cite{Ayach_Spatially_14, Alkhateeb_channel_2014}, has been receiving increasing interest. In the latter, the number of RF chains at the transmitter and receiver is drastically smaller than the number of antennas. Moreover, both the precoding and combining are done in two stages, digital and analog. However, many fundamental questions surrounding the latter architecture have to be answered, namely,  how to estimate the large mmWave channel, and design the digital / analog precoders and combiners. Though an algorithm was proposed in~\cite{Alkhateeb_channel_2014} for that purpose, the latter requires knowledge of the number of propagation paths (i.e. propagation environment), it exhibits relatively elevated complexity, and builds an estimate of the entire channel, that is then used to design the precoding / combining. 

Rather than estimating the entire channel, our proposed Krylov subspace method exploits the reciprocity of the channel in TDD MIMO systems, and directly estimates the right (resp. left) singular subspace at the transmitter (resp. receiver) - required for optimal transmission. Moreover, we propose an algorithm for subspace decomposition, whereby each of the estimated subspaces is approximated by a cascade of the digital and analog precoder, while satisfying the constraints of the hybrid architecture. We underline the fact that this proposed approach is perfectly applicable to conventional MIMO systems, i.e. fully digital, as well.  We note that some parts of this works are based on~\cite{Ghauch_BSE_journ}, therefore some discussions / derivations / proofs / algorithms are omitted here.

In the following, we use bold upper-case letters to denote matrices, and bold lower-case denote vectors. Furthermore, for a given matrix $\pmb{A}$, $[\pmb{A}]_{i:j}$ denotes the matrix formed by taking columns $i$ to $j$, of $\pmb{A}$,  $\Vert \pmb{A} \Vert_F^2$ its Frobenius norm, $|\pmb{A}|$ its determinant, $\pmb{A}^\dagger$ its conjugate transpose. $ [\bA]_{i,j} = a_{i,j}$ denotes element $(i,j)$ in a matrix $\bA$, and $[\ba]_i$ element $i$ in a vectors $\ba$. While $\bI_n$ denotes the $n \times n$ identity matrix,  $\pmb{1}_n$ denotes the $n \times 1$ vector of ones. Finally, we let $\lbrace n \rbrace \triangleq \lbrace 1, ..., n \rbrace $, and $\calS_{p,q} = \left\lbrace \bX \in \bbC^{p \times q} \ | \ \ \vert \bX_{ij} \vert = 1/\sqrt{p} \ , \ \forall (i,k) \in \lbrace p \rbrace \times \lbrace q \rbrace \right\rbrace $.

\section{System Model}
Assume a single user MIMO system with $M$ and $N$ transmit antennas at the BS and MS, respectively,  where each is equipped with $r$ RF chains, and sends $d$ independent data streams ($d \leq r \leq \min(M,N)$). The downlink (DL) received signal, after filtering, is given by, 
\begin{align}\label{eq:sigmod}
\tilde{\bx}^{(r)} &= \bU^\dagger \bW^\dagger \bH \bF \bG \bx^{(t)} + \bU^\dagger \bW^\dagger \bn^{(r)} 
\end{align}
where  $\bH \in \bbC^{N \times M }$ is the complex channel - assumed to be slowly block-fading,  $\bF \in \mathbb{C}^{M \times r}$ is the analog precoder, $\bG \in \mathbb{C}^{r \times d}$ the digital precoder, $\bx^{(t)}$ is the $d$-dimensional transmit signal with covariance matrix $E[\bx^{(t)} \bx^{{(t)}^\dagger}] = (P_s/d) \bI_d$ and $\bn^{(r)}$ is the AWGN noise at the receiver, with $E[\bn^{(r)} \bn^{{(r)}^\dagger}] = \sigma_r^2 \bI_N $. Similarly,  $\bW \in \mathbb{C}^{N \times r}$ and $\bU \in \mathbb{C}^{r \times d}$ are the analog and digital combiner, respectively.
In addition to requiring both the analog precoder and combiner to have constant modulus elements, i.e., $\bF \in \calS_{M,r}$ and $\bW \in \calS_{N,r}$ (since the latter represent phase shifters), a total power constraint must still satisfied, i.e., $\Vert \bF \bG \Vert \leq \rho^2 d$ (where we assume that $\rho = 1$ w.l.o.g.).\footnote{Similarly, exploiting channel reciprocity, the uplink received signal is given by  $\tilde{\bx}^{(t)} = \bG^\dagger \bF^\dagger \bH^\dagger \bW \bU \bx^{(r)} + \bn^{(t)}$  where $\by^{(t)}$ is the $M$-dimensional signal at the transmitter and $\bn^{(t)}$ is the AWGN noise at the transmitter, such that $E[\bn^{(t)} \bn^{{(t)}^\dagger}] = \sigma_t^2 \bI_N $} 
We also assume a TDD system where channel reciprocity holds, and  denote the SVD of $\bH$ as,
\begin{align} \label{eq:svdH}
\bH =   
\begin{bmatrix}
\bPhi_1 , & \bPhi_2
\end{bmatrix}
\begin{bmatrix}
\bSig_1  & \bo \\
\bo & \bSig_2
\end{bmatrix}
\begin{bmatrix}
\bGam_1^\dagger \\
\bGam_2^\dagger 
\end{bmatrix}
= \bPhi_1 \bSig_1 \bGam_1^\dagger + \bPhi_2 \bSig_2 \bGam_2^\dagger
\end{align}
where $\bGam_1 \in \bbC^{M \times d}$ and $\bPhi_1 \in \bbC^{N \times d} $ are unitary, and $\bSig_1 \in \bbC^{d \times d} $ is diagonal with the $d$-largest singular values of $\bH$ (recall that $\bGam_1^\dagger \bGam_2 = \bo$ and $\bPhi_1^\dagger \bPhi_2 = \bo$ ). 
In view of clarifying the aim of our work, we present the following intuitive result. 
\begin{proposition} \label{prop:opttx}
Given the signal model in~\eqref{eq:sigmod}, the optimal analog and digital precoder / combiner that maximize the user rate are such that $ \bF \bG = \bGam_1  $ and $ \bW \bU = \bPhi_1  $ (assuming waterfilling power allocation is performed over the resulting effective channel).
\end{proposition}
Though the latter result is expected, it is reminiscent of the well-known optimal transmission strategy for classical MIMO, where the transmitter uses right singular vectors, $\bGam_1$, for precoding, and receiver uses the left singular vectors, $\bPhi_1$, for combining: the above proposition suggests that this structure still maximizes the user rate in the hybrid architecture, provided one is able to approximate  $ \bF \bG$ by $ \bGam_1 $, and $ \bW \bU$ by $\bPhi_1 $ (and assuming that waterfilling is employed). Since no a priori CSI is assumed to be available at neither the transmitter nor the receiver, our aim is firstly to propose an algorithm to estimate $\bGam_1$ at the transmitter, i.e. $\tbGam_1$, and $\bPhi_1$ at the receiver, i.e. $\tbPhi_1$. This done, we shed light on the problem of subspace decomposition, and present an algorithm for approximating the estimated subspaces, $\tbGam_1$ by $\bF \bG $ and $\tbPhi_1$ by $\bW \bU $. We describe our scheme in the context of conventional MIMO systems, i.e. fully digital, and later extend it to the hybrid architecture.

\section{Eigenvalue algorithms and Subspace Estimation} \label{sec:bsemimo}
With this mind, the aim of subspace estimation algorithms is to obtain \emph{$\tbGam_1$ at the transmitter} (keeping in mind that $\bGam_1$ is nothing but the dominant eigenvectors of $\ \bH^\dagger \bH$), and \emph{$\tbPhi_1$ at the receiver}. We note that eigenvalue algorithms such as the Power Method or Subspace Iteration, well known from numerical analysis, were used in~\cite{Dahl_blind_04} for that same purpose. In this work we resort to \emph{Krylov subspace methods}, to achieve the latter goal. One such method is the well-known Arnoldi Iteration (the variant we use here is detailed in~\cite{Saad_Numerical_11}) whereby one starts with a random vector $\bq_1$, and recursively builds $\bQ_m \triangleq [\bq_1 , ....,  \bq_m] \in \bbC^{M \times m} \ (m \leq M) $ such that $$\bQ_m^\dagger (\bH^\dagger \bH) \bQ_m = \bT_m, \ \bQ_m^\dagger \bQ_m = \bI_m $$
where $\bT_m \in \bbC^{m \times m} $ is an upper Hessenberg matrix, and the resulting $\bQ_m$ is an orthonormal basis for the Krylov subspace in question. Consequently, the eigenpairs of $\bT_m$ are eigenpairs of $\bH^\dagger \bH$, and the desired subspace $\bGam_1$ can be computed by finding the eigenpairs of $\bT_m$ - which can be found efficiently. 

A more careful examination quickly reveals that implementing the latter method in a distributed way requires the transmitter to have the sequence $\lbrace \bH^\dagger \bH \bq_1, \cdots, \bH^\dagger \bH \bq_m \rbrace $. Without any prior channel knowledge, this can be accomplished using the echoing mechanism that was employed in~\cite{Dahl_blind_04}, whereby the transmitter sends $\bq_l$ in the DL, and it is echoed back by the receiver using Amplify-and-Forward (A-F), as follows, 
\begin{align} \label{eq:echo}
//DL: \ \ \bs_l &= \bH \bq_l + \ww{r}{l} \nonumber   \\
//UL: \ \ \bp_l &= \bH^\dagger \bs_l + \ww{t}{l} = \bH^\dagger \bH \bq_l + \bH^\dagger \ww{r}{l} + \ww{t}{l}   
\end{align}   
After the echoing phase, the transmitter has a noisy estimate, $\bp_l$, of $\bH^\dagger \bH \bq_l$, as seen from~\eqref{eq:echo}. We note that incorporating noise, i.e., $\ww{r}{l}$ and $\ww{t}{l}$ in the algorithm formulation, allows us to extend the original formulation of the Arnoldi Iteration, to account for external distortion, and provide bounds on the estimation error (further details are provided in~\cite{Ghauch_BSE_journ}, where we derive bounds on the estimation error of the subspaces in question). Steps 2.a - 3.a follow the conventional Arnoldi iteration. Finally, computing the estimate of $\bGam_1$ (steps 4.a - 4.c) follows immediately from the fact that the eigenvectors of $\bT_m$, at the output of the Arnoldi iteration, approximate the Ritz eigenvectors of $\bH^\dagger \bH$~\cite{Saad_Numerical_11}. The above steps are summarized in the Subspace Estimation using Arnoldi Iteration (SE-ARN) procedure below.
\begin{algorithm} 
\begin{algorithmic}
\State \textbf{Subspace Estimation using Arnoldi Iteration (SE-ARN)}
\Procedure{$\tbGam_1=$ SE-ARN }{$\bH$, $d$}  
\State Set $m$ ($m \leq M$);  Random unit-norm $\bq$;  $\bQ = [\bq_1]$
\For {$l = 1, 2, ..., m $}
	\State // \emph{transmitter-initiated echoing: estimate $\bH^\dagger \bH \bq_l$}
	\State 1.a \hspace{.1cm} $\bs_l = \bH \bq_l + \ww{r}{l}  $
	\State 1.b \hspace{.1cm} $\bp_l = \bH^\dagger \bs_l + \ww{t}{l} $
	\State // \emph{Gram-Schmidt orthogonalization}
	\State 2.a \hspace{.1cm} $t_{m,l} = \bq_m^\dagger \bp_l \hspace{.7cm}  , \forall \ m = 1, \dots, l $
	\State 2.b \hspace{.1cm} $\br_l = \bp_l - \sum_{m=1}^l \bq_m t_{m,l} $
	\State 2.c \hspace{.1cm} $t_{l+1,l} = \Vert \br_l \Vert_2 $
	\State // \emph{Update $\bQ$} 
	\State 3.a \hspace{.1cm} $\bQ = [ \bQ ,  \  \bq_{l+1} = \br_l/t_{l+1, l} ]$
\EndFor
\State // \emph{Compute $\tbGam_1$} 
\State  4.a \hspace{.1cm} $ \bT_m = \tbTh \tilde{\bLam} \tbTh^{-1}$ 
\State  4.b \hspace{.1cm} $\tbGam_1 = \bQ_m \tbTh_{1:d}$ 
\State  4.c \hspace{.1cm} $\tbGam_1 = \ \textrm{qr} (\tbGam_1)$ 
\EndProcedure
\end{algorithmic}
\end{algorithm}
 \section{Hybrid precoding for mmWave MIMO systems}
In this section we extend the previous framework to fit the hybrid architecture, and highlight the major challenges. We first start by presenting some preliminaries that will later be used throughout this section. 

\subsection{Preliminaries: Subspace Decomposition}
We assume that $d$ of the $r$ available RF chains are used, i.e., $\bF \in \bbC^{M \times d}$ and $\bG \in \bbC^{d \times d}$ (more on that, later in this section). In conventional MIMO systems, once the estimates, $\tbGam_1$ and $\tbPhi_1$, are obtained they can immediately be used as transmit and receive filters, respectively. However, in the case of the hybrid architecture, as Proposition~\ref{prop:opttx} suggests, $\tbGam_1$ needs to be expressed as $\bF \bG$ (moreover $\tbPhi_1$ needs to be expressed as $\bW \bU$, but we restrict the discussion to the transmitter, for brevity), while satisfying both the maximum power and hardware constraints. Using the Frobenius norm a distance metric - a rather simple engineering heuristic, we formulate the subspace decomposition (SD) problem as follows, 
\begin{align} \label{opt:qp}
\begin{cases}
               \underset{\bF, \ \bG}{\min} \ \ h_0(\bF, \bG) = \Vert \tbGam_1 - \bF \bG  \Vert_F^2   \\
               \st \ \ h_1(\bF, \bG) = \Vert \bF \bG \Vert_F^2 \leq d \\
               \hspace{.8cm} \bF \in \calS_{M,d}
            \end{cases}
\end{align}

\subsubsection{Block Coordinate Descent for Subspace Decomposition}
\begin{figure}
	\center
	\includegraphics[width=10cm, height=6.5cm]{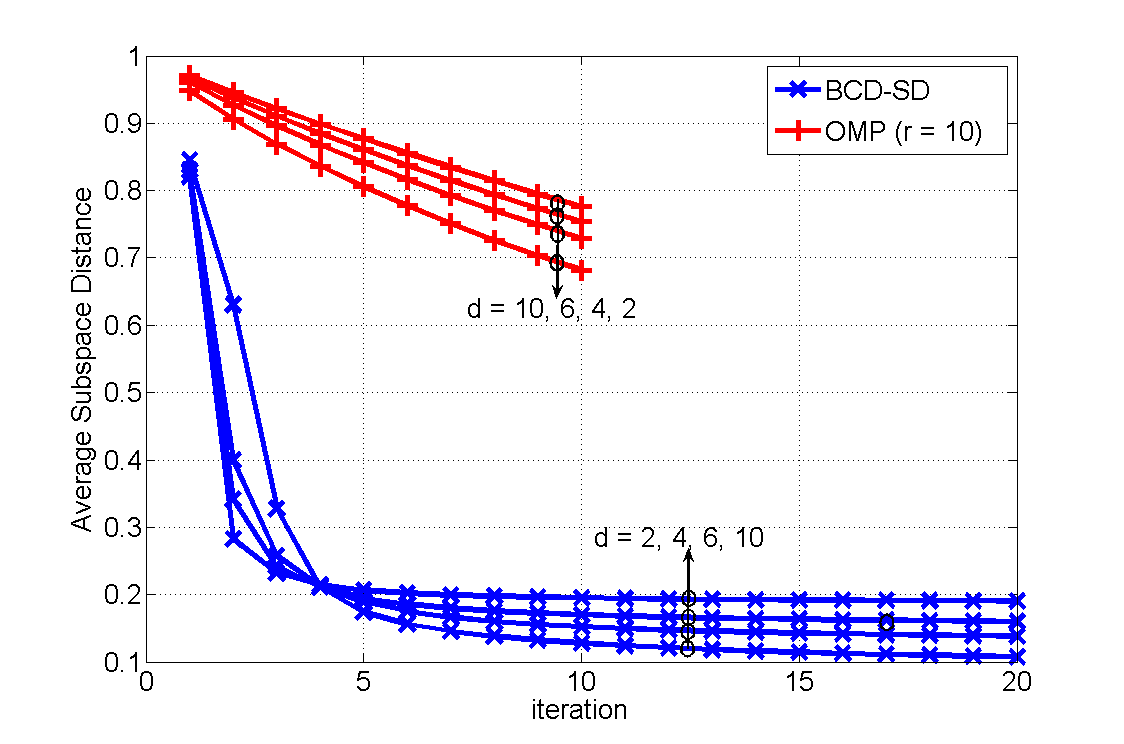}
	\caption{ Average subspace distance $\Vert \tbGam_1 - \bF \bG \Vert_F^2$ }
	\label{fig:dcmp}
\end{figure}

Due to the coupled nature of~\eqref{opt:qp}, Block Coordinate Decent (BCD) stands out as an attractive approach, whereby $\bF$ and $\bG$ are iteratively updated,  such that the sequence $\lbrace h_0(\bF_k , \bG_k ) \rbrace_k $ is non-increasing. We will subsequently show that the updates resulting from the BCD method implicitly enforce a power constraint (consequently, the latter can be dropped from~\eqref{opt:qp}). Relaxing the hardware constraint on $\bF$, we first fix $\bG$ and optimize $\bF$, and vice versa. Note that that resulting sub-problems problems are instances of a non-homogeneous convex QCQP that can be solved using standard Lagrangian techniques, to yield the following solutions, 
\begin{align} \label{eq:Fopt}
&\bF_{k+1} =  \tbGam_1 \bG_k^\dagger (\bG_k \bG_k^\dagger)^{-1} 
\end{align}
\begin{align} \label{eq:Gopt}
&\bG_{k+1} =  (\bF_{k+1}^\dagger \bF_{k+1})^{-1} \bF_{k+1}^\dagger \tbGam_1 
\end{align} 
Note that our earlier assumption that only $d$ RF chains are used, ensures that $(\bG_k \bG_k^\dagger)$ in~\eqref{eq:Fopt} is invertible. Moreover, using simple manipulations (and assuming w.l.o.g. that $\Vert \tbGam_1 \Vert_F^2 = 1$) it can be shown that $$ \Vert \bF_{k+1} \bG_{k+1} \Vert_F^2 \leq d , \ \forall k$$ implying that the power constraint is indeed enforced. Recall that $\bF_{k+1}$ in~\eqref{eq:Fopt} does not necessarily satisfy the hardware constraint. It can be shown that its (unique) Euclidean projection on the set $\calS_{M,d}$, i.e.,$$\tbF_{k+1} \triangleq \Pi_{\calS}[\bF_{k+1}] = \underset{\bU \in \calS_{M,d} }{\argmin} \ \Vert \bU - \bF_{k+1} \Vert_F^2 $$ is given by $[\tbF_{k+1}]_{m,n} = (1/ \sqrt{M}) e^{j\phi_{m,n}} , \ \forall (m,n) $, where $ \phi_{m,n} \triangleq  \arg ( [\bF_{k+1}]_{m,n} )$. The corresponding algorithm, Block Coordinate Descent for Subspace Decomposition (BCD-SD) is shown below. Note that the latter projection makes convergence claims extremely difficult to make. 
\begin{algorithm} 
\begin{algorithmic}
\State \textbf{Block Coordiate Descent for Subspace Decomposition (BCD-SD)}
\Procedure{[$\bF$, $\bG$] $=$ BCD-SD }{$\tbGam_1$, $\rho$}  
\State Start with arbitrary $\bG_0$
\For {$k = 0, 1, 2, ... $}
	\State $\bF_{k+1} \leftarrow  \Pi_{\calS}\left[ \tbGam_1 \bG_k^\dagger (\bG_k \bG_k^\dagger)^{-1} \right]  $ 
	\State $\bG_{k+1} \leftarrow (\bF_{k+1}^\dagger \bF_{k+1})^{-1} \bF_{k+1}^\dagger \tbGam_1 $ 
\EndFor
\EndProcedure
\end{algorithmic}
\end{algorithm}
We note that the authors in~\cite{Ayach_Spatially_14} formulated the same problem as~\eqref{opt:qp} after a series of approximations to the mutual information, and proposed a variation on the well-known Orthogonal Matching Pursuit (OMP), whereby the columns of $\bF$ are iteratively recovered in a greedy manner. We thus compare its average performance with our proposed method, for a case where $\tbGam_1 \in \bbC^{M \times d} $ is such that $M=64, r = 10$ (for several values of $d$). The reason for the massive performance gap in Fig.~\ref{fig:dcmp} is that our proposed method attempts to find a locally optimal solution to~\eqref{opt:qp} (though this cannot be shown due to the projection step). Moreover, OMP is halted after $r$ iterations, since it recovers the columns of $\bF$ one at at time, whereas our proposed method runs until reaching a stable point. 

\subsubsection{Beamforming case} \label{sec:dcmp}
The case where $d=1$ in~\eqref{opt:qp} is of particular importance. Recall that echoing received vectors is the mechanism at the heart of our approach. For the hybrid architecture this implies that both transmitter and receiver need to be able to approximate any digital beamforming vector $\bq_l$, by $\bF \bG$, where $\bff$ is a vector and $g$ is a scalar. When $d=1$, it can be shown that~\eqref{opt:qp} reduces to the problem below.
\begin{lemma} \label{lem:bfsol}
Consider single dimension SD problem,
\begin{align} \label{opt:bfb}
\begin{cases}
               \underset{\bff, \ g}{\min} \ h_o(\bff, \ g) = \Vert \bff \Vert_2^2 \ g^2 -  2g\Re(\bff^\dagger \tbgam_1 )  \\
               \st  \ [\bff]_i = 1 /\sqrt{M} \ e^{j\phi_i} , \forall i  
            \end{cases}
\end{align}
where $g \in \mathbb{R}_+ $ and $[\tbgam_1]_i  = r_i e^{j \theta_i}$. Then the problem admits a globally optimum solution given by, $[\bff^\star]_i = 1 /\sqrt{M} \ e^{j\theta_i}, \forall i $ and $g^\star = \Vert \tbgam_1 \Vert_1 / \sqrt{M}$
\end{lemma}
\begin{IEEEproof} Refer to~\cite{Ghauch_BSE_journ} for proof \end{IEEEproof}
Moreover, the approximation error $\be \triangleq \tbgam_1 - \bff g$ is such that, 
\begin{align} \label{eq:dcmperr}
\vert [\be]_i \vert = \vert r_i - \Vert \tbgam_1 \Vert_1 /M \vert e^{j \theta_i}, \ \forall i \in \lbrace M \rbrace.
\end{align}

\subsection{Echoing in Hybrid Architecture}
\subsubsection{Motivation}
For the sake of simplicity, we neglect noise from our formulations, and focus on other sources of distortion. Recall that the proposed scheme requires $\lbrace \bH^\dagger \bH \bq_l \rbrace_{l=1}^m$ at the transmitter. Though this can be easily done in conventional MIMO systems (using the transmitter-initiated echoing mechanism in~\eqref{eq:echo}), the A-F step required by the receiver is not possible in the hybrid architecture.\footnote{Recall that digitally processing the baseband signal is only possible after the application of the analog precoder / combiner (and possibly the digital precoder / combiner)~\cite{Ayach_Spatially_14}.} With this in mind, one can naively attempt to emulate the A-F step in transmitter-initiated echoing, described in~\eqref{eq:echo}, as follows: decompose  $\bq_l$ at the transmitter, into $\tbf_l \tilde{g}_l$, i.e. $\bq_l = \tbf_l \tilde{g}_l + \be_l$, and send $\tbf_l \tilde{g}_l$ over the DL; processes the received signal in the downlink, with the analog combiner, i.e., $\bs_l = \bW_l^\dagger(\bH \tbf_l \tilde{g}_l )$; apply same filter  to process the transmit signal in the UL, i.e., $\bW_l \bs_l $. Finally, the received signal the the transmit antennas is processed with the analog precoder $\bF_l$. The resulting signal at the transmitter is, 
\begin{align} \label{eq:echowrong}
\bp_l = \bF_l^\dagger \bH^\dagger \bW_l \bW_l^\dagger \bH (\bq_l - \be_l ) 
\end{align} 
It is clear from~\eqref{eq:echowrong} that $\bp_l$ is no longer a ``good'' estimate of $\bH^\dagger \bH \bq_l$. Firstly, the fact that the signals at the receiver (resp. transmitter) need to be processed with the analog combiner $\bW_l$ (resp. precoder $\bF_l$) implies that the desired estimate of $\bH^\dagger \bH \bq_l$ is distorted. Moreover, the application of $\bF_l \in \bbC^{M \times r}$ in~\eqref{eq:echowrong} implies that the dimension of the estimate is reduced from $M$ to $r$. We dub such distortions \emph{Analog-Processing Impairments (API)}. In addition, the estimate of $\bH^\dagger \bH \bq_l$ is further distorted by the decomposition error, $\be_l$, emanating  from decomposing $\bq_l$ at the transmitter (which we refer to a \emph{Decomposition-Induced Distortion (DID)}) The above impairments are a by-product  of the constraints imposed by the hybrid architecture, and will individually be investigated and addressed.

\subsubsection{Cancellation of Analog-Processing impairments} \label{sec:did}
Our proposed method for mitigating analog-processing impairments (API) relies on the simple idea of taking multiple measurements at both transmitter and receiver, using carefully chosen analog precoders / combiners, such that $\bW_l\bW_l^\dagger$ and $\bF_l \bF_l^\dagger$  approximate an identity matrix. 

In the DL, $\bq_l$ is approximated by $ \tbf_l \tilde{g}_l$, and  $\tbf_l \tilde{g}_l$ is sent over the DL channel\footnote{Instead of using only one RF chain to send $\tbf_l \tilde{g}_l$ over the DL, we use all the available $d$ RF chains, 
thereby resulting in an array gain factor of $d$. We also make use of this observation in the UL sounding.}, 
$K_r$ times (where $K_r = N/r$), each linearly processed with an analog combiner $\lbrace \bW_{l,k} \in \bbC^{N \times r} \rbrace_{k=1}^{K_r}$, to obtain the digital samples $\lbrace \bs_{l,k} \rbrace_{k=1}^{K_r}$. Moreover, the analog combiners  are taken from the columns of a Discrete Fourier Transform (DFT) matrix, i.e, 
\begin{align} \label{eq:dft1}
&[\bW_{l,1}, ..., \bW_{l,K_r}] = \bD_r, 
\end{align}
where $\bD_r \in \bbC^{N \times N}$ is a normalized $N \times N$ DFT matrix. The same analog combiners, $\lbrace \bW_{l,k} \rbrace_{k}$, are used to linearly combine $\lbrace \bs_{l,k} \rbrace_{k}$, to form $\tbs_l$ . The above steps are summarized in the Repetition-Aided (RAID) Echoing procedure below. Combining the above equations, we rewrite $\tbs_l$ as, 
\begin{align} \label{eq:tbs}
\tbs_l = \left( \sum_{k=1}^{K_r} \bW_{l,k} \bW_{l,k}^\dagger \right) \bH (d \tbf_l \tilde{g}_l) = d \bH  \tbf_l \tilde{g}_l 
\end{align}
where equality follows from the fact that $\lbrace \bW_{l,k}  \rbrace_k$ are columns of a DFT matrices. Note that \emph{the effect of processing the received signal with the analog combiner has been completely suppressed}.
\begin{algorithm} 
\begin{algorithmic}
\State \textbf{Repetition-Aided (RAID) echoing}
\State // \emph{DL phase}
\State \hspace{.5cm} $\bq_l = \tbf_l \tilde{g}_l + \be^{(t)}_l$
\State \hspace{.5cm} $\bs_{l,k} = \bW_{l,k}^\dagger \bH (d \tbf_l \tilde{g}_l), \ \forall k \in \lbrace K_r \triangleq N/r \rbrace$
\State \hspace{.5cm} $\tbs_l = \sum_{k=1}^{K_r} \bW_{l,k} \bs_{l,k}$
\State // \emph{UL phase}
\State \hspace{.5cm} $\tbs_l = \tbw_l \tilde{u}_l + \be^{(r)}_l $
\State \hspace{.5cm} $\bz_{l,m} = \bF_{l,m}^\dagger \bH^\dagger (d \tbw_l \tilde{u}_l ), \ \forall m \in \lbrace K_t \triangleq M/r \rbrace  $
\State \hspace{.5cm} $\bp_l = \sum_{m=1}^{K_t} \bF_{l,m} \bz_{l,m}   $
\end{algorithmic}
\end{algorithm}
The exact same process is used in the UL:  $\tbs_l$ is first decomposed into $\tbw_l \tilde{u}_l$, i.e. $\tbs_l = \tbw_l \tilde{u} + \be_l^{(r)}$ , $d$ RF chains are used to send it over the UL, $K_t$ times (where $K_t = M/r$), and each observation is linearly processed with an analog precoder $\lbrace \bF_{l,m} \in \bbC^{M \times r} \rbrace_{m=1}^{K_t}$, where  the latter is taken from the columns of a DFT matrix. The process for the UL is summarized in the RAID echoing procedure. We combine the latter steps to rewrite $\bp_l$ as, 
\begin{align} \label{eq:bp}
\bp_l &= \left( \sum_{m=1}^{K_t} \bF_{l,m} \bF_{l,m}^\dagger \right) \bH^\dagger (d \tbw_l \tilde{u}_l ) = d \bH^\dagger  \tbw_l \tilde{u}_l 
\end{align}
Thus, the output of the RAID procedure is as follows,
\begin{align}
\bp_l &= d \bH^\dagger \tbw_l \tilde{u}_l = d \bH^\dagger (\tbs_l - \be^{(t)}_l ) =  d \bH^\dagger (d \bH \tbf_l \tilde{g}_l - \be^{(t)}_l )   \nonumber   \\
&= d^2 \bH^\dagger \bH \bq_l - d^2 \bH^\dagger \bH \be^{(t)}_l - d \bH^\dagger \be^{(r)}_l   \label{eq:pl}
\end{align}
where $\be_l^{(t)}$ (resp. $\be_l^{(r)}$) is the transmitter-side DID (resp. receiver-side DID) resulting from decomposing the digital transmitted signal at the transmitter (resp. receiver). It is quite insightful to compare $\bp_l$ in  the latter equation with~\eqref{eq:echowrong}. We can clearly see that \emph{impairments originating from processing the received signals with both $\bW_l$ and $\bF_l$, have completely been suppressed}: in~\eqref{eq:pl}, $\bp_l$ indeed is the desired estimate, i.e., $\bH^\dagger \bH \bq_l$, corrupted by distortions. Note that employing \emph{this process reduces the hybrid architecture into a conventional MIMO channel}: any transmitted vector in the DL, $(\tbf_l \tilde{g}_l) $, can be received in a ``MIMO-like'' fashion, as seen from~\eqref{eq:tbs}, at a cost of $K_r$ channel uses (the same holds for the UL, as seen from~\eqref{eq:bp} ).  


\subsubsection{Decomposition-Induced Distortion (DID)}
We investigate the effect of transmitter-side DID, $\be_l^{(t)}$, and receiver-side DID, $\be_l^{(r)}$, that distort $\bp_l$, at the output of the RAID procedure in~\eqref{eq:pl}. It can be easily verified that $\be_l^{(t)}$ only distorts the magnitude of $\bH^\dagger \bH \bq_l$, not its phase, and consequently its effect is minimal and can be neglected. Since this claim cannot be made for the receiver-side DID, $\be_l^{(r)}$, we provide a mechanism for mitigating the latter, however, at the cost of additional communication overhead. The details of the latter technique are further elaborated in~\cite{Ghauch_BSE_journ}, but omitted here due to space limitations.

\subsection{Proposed Algorithm}
We now formulate our algorithm for Subspace Estimation and Decomposition (SED) in the mmWave architecture (shown in Algorithm~\ref{alg:bsehyb}): estimates of the right / left singular subspaces, $\tbGam_1$ and $\tbPhi_1$, can be obtained by using the SE-ARN procedure (Sect.~\ref{sec:bsemimo}), keeping in mind that the echoing phase (Steps 1.a and 1.b) is now replaced by the RAID echoing procedure (Sect.~\ref{sec:did}) . Then, the multi-dimensional subspace decomposition procedure, BCD-SD in Sect.~\ref{sec:dcmp}, is then used to approximate each of the estimated singular spaces, by a cascade of analog and digital precoder / combiner. Note that the total communication overhead required by the algorithm is $\Omega = 2m (M+N)/r  $ channel uses ($m$ being the number of iterations for the SE-ARN). Moreover, recall that the lack of statistical models for mmWave channels, and the fact that MMSE channel estimates cannot be obtained in a hybrid analog-digital MIMO system, make it difficult to analyze the effect of channel estimation errors. 
\begin{algorithm} 
\caption{Subspace Estimation and Decomposition (SED) for Hybrid Architecture} \label{alg:bsehyb}
\begin{algorithmic}
\State // \emph{Estimate $\tbGam_1$ and $\tbPhi_1$ }
\State $\tbGam_1 =$ SE-ARN ($\bH$, $d$) 
\State $\tbPhi_1 =$ SE-ARN ($\bH^\dagger$, $d$) 
\State // \emph{Decompose $\tbGam_1$ and $\tbPhi_1$ }
\State [$\bF$, $\bG$ ] = BCD-SD ($\tbGam_1$, $\rho$) 
\State [$\bW$, $\bU$ ] = BCD-SD ($\tbPhi_1$, $\rho$) 
\end{algorithmic}
\end{algorithm}

\section{Numerical Results}
Though our approach is not restricted to any particular channel model, for our numerical evaluations, we adopt the prevalent channel model in the mmWave  literature, where  only $L$ scatterers are assumed to contribute to the received signal (an inherent property of their poor scattering nature), 
\begin{align} \label{eq:chann}
\bH = \sqrt{\frac{M N}{L	}} \sum_{i=1}^L \beta_i  \ \ba_r(\chi_i^{(r)}) \ba_t^\dagger(\chi_i^{(t)} )
\end{align}
where $\chi_i^{(r)}  $ and $\chi_i^{(t)}$ are angles of arrival at the MS, and angles of departure at the BS (AoA / AoD) of the $i^{th}$ path,  respectively (both assumed to be uniform over $[-\pi/2, \ \pi/2]$), $\beta_i$ is the complex gain of the $i^{th}$ path such that $\beta_i \sim \mathcal{CN}(0, 1), \  \forall i $. Finally, $\ba_r(\chi_i^{(r)})$ and $\ba_t(\chi_i^{(t)})$ are the array response vectors at both the MS and BS, respectively (assumed to be uniform linear arrays). We assume that the number of RF chains scales with the number of antennas, e.g.,  $M/r = 8$ and $N/r = 4$. Though it remains to be seen whether it is achievable, we use the following user rate as a our metric~\cite{Baum_sounding_11},
\small
\begin{align*}
 R = \log_2 \left| \bI_d + \frac{P_s}{d \sigma_r^2} \bU^\dagger \bW^\dagger \tbH \bF \bG \bG^\dagger \bF^\dagger \tbH^\dagger \bW  \bU (\bU^\dagger \bW^\dagger \bW \bU)^{-1} \right|
\end{align*}
\normalsize
 where $\tbH$ is the channel estimate resulting from our proposed method. Note that an algorithm for mmWave MIMO channel estimation was proposed in~\cite{Alkhateeb_channel_2014}. However, since many of its underlying details are not present in the paper, we opt to use a simple \emph{Independent Sounding} sounding scheme: the analog precoder and combiner are first selected by exhaustively sounding DFT codebooks at both transmitter and receiver, then the digital precoder and combiner are chosen as right and left singular vectors of the effective channel estimate. We adjust the number of iterations for our scheme, $m$, such that the resulting communication overhead is similar to that of the benchmark scheme. In addition, we use a \emph{perfect CSI, fully digital case} (i.e. the capacity of equivalent MIMO channel with perfect CSIT / CSIR) as an upper bound. All curves are averaged over $500$ channel realizations.

\begin{figure}
	\center
	\includegraphics[width=9.5cm, height=7cm]{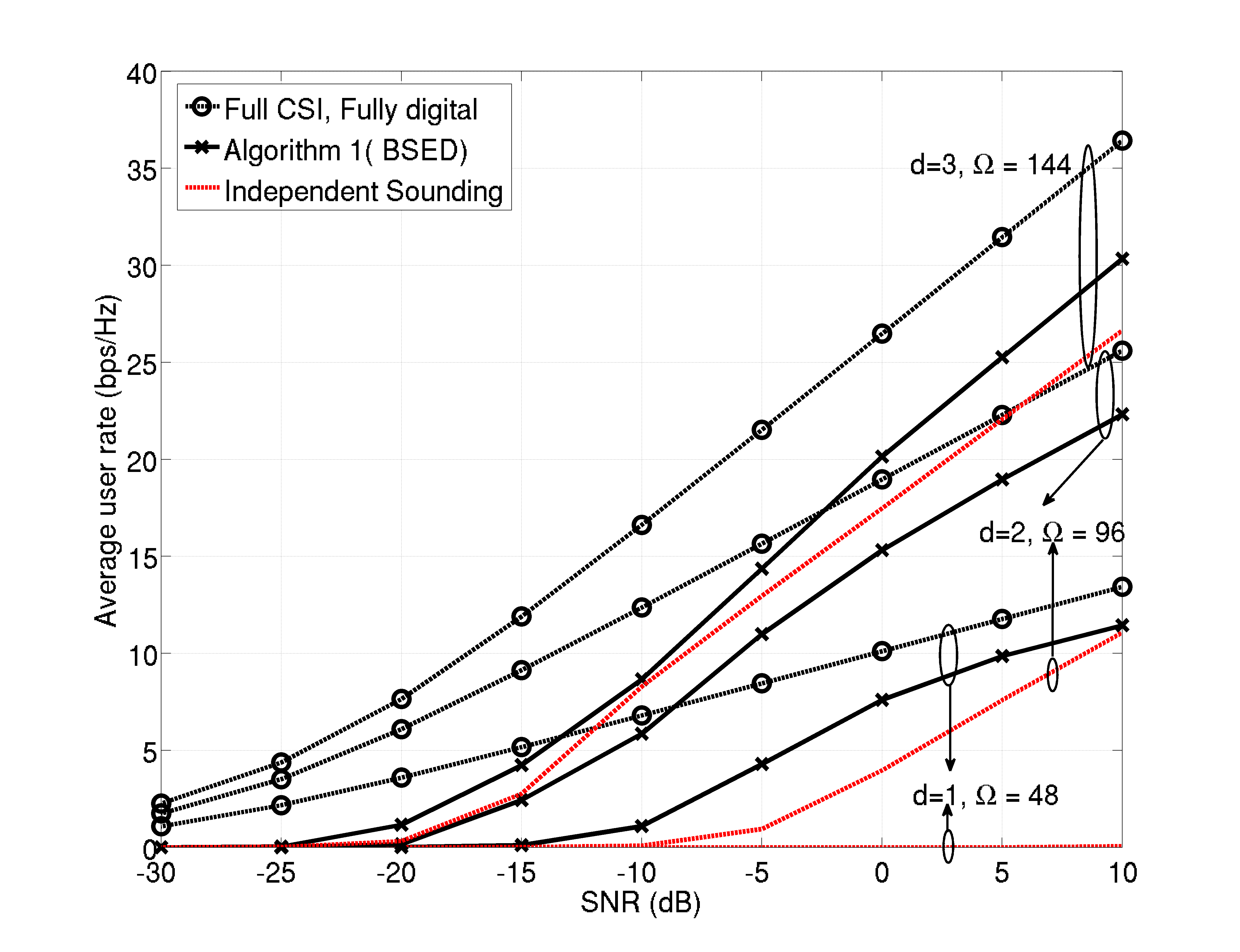}
	\caption{Average user rate of proposed schemes over SCM channels ($M=64, N=32, m = 2d$) }
	\label{fig:sr2}
	\center
	\includegraphics[width=9.5cm, height=7cm]{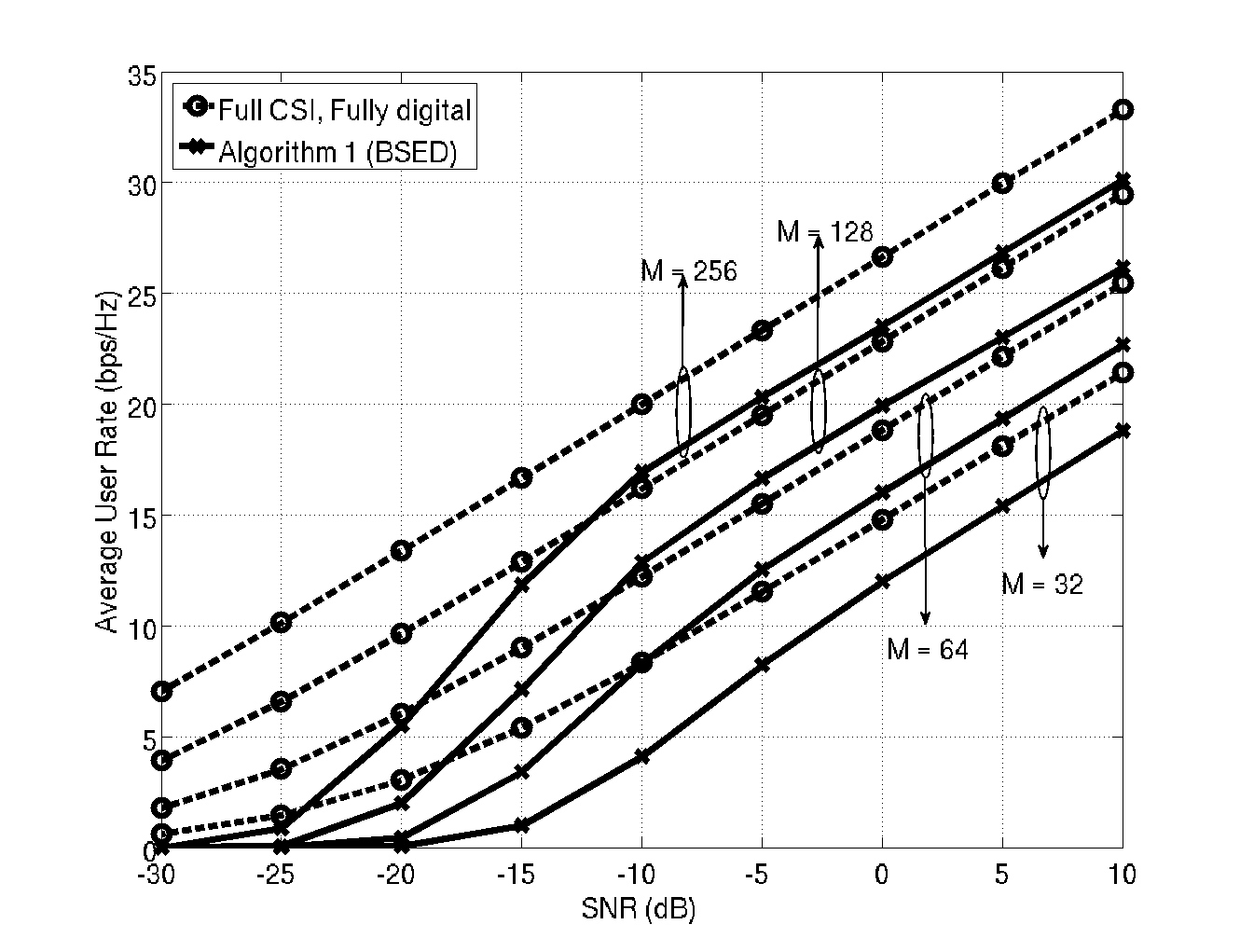}
	\caption{Average user rate for different $M,N$ ($N=M/2, d=2, L=4, m =6$) }
	\label{fig:sr3}
\end{figure}

In view of having a more realistic performance evaluation, we adopt the Spatial Channel Model (SCM) detailed in~\cite{3GPP_tr25996}, and modify its parameters to emulate mmWave channels described above (where a small value of $\Omega$ is desired).
Fig.~\ref{fig:sr2} shows the user rate of such a system, with $M=64, N=32, m = 2d$, for several values of $d$ (each resulting in different values for $\Omega$). We can clearly see that our scheme yields a significantly high throughput in this realistic simulation setting (especially for $d=3$), while still keeping the overhead at a relatively low level. Interestingly, we can see that the benchmark scheme offers a surprisingly poor performance, except for the case when $d=3$ (since in this case, the receiver codebook consists of the entire DFT matrix). This does suggest that the performance of the independent sounding scheme is highly unstable, and very much dependent on the size of the codebooks. 
We next investigate the scalability of our proposed scheme, by scaling up $M$ and $N$ (assuming $N= M/2$ for simplicity), while keeping everything else fixed, i.e. $d=2 , m=6$, and consequently $\Omega = 144$. Fig~\ref{fig:sr3} clearly shows that the algorithm is able to harness the significant array gain inherent to large antenna systems, while keeping the overhead the same. Though the performance might not be good enough to offset the overhead, for the $32 \times 16$ case, it surely does for the $256 \times 128$ (the key to this impressive result is to have $M/r$ and $N/r$ fixed, as $M,N$ increase). Moreover, it is hard to ascertain whether the low-SNR gap (between the ideal case and Algorithm~\ref{alg:bsehyb}) is due to our proposed estimation method, or inherent to the problem of mmWave MIMO channel estimation.

Consequently, our results indeed suggest that \emph{the performance achieved by conventional MIMO systems can still be maintained in the hybrid architecture, with a drastically reduced number of RF chains} ($\sim 4$ to $\sim 8$ times less), thereby resulting in massive savings in terms of cost and power consumption.

\section{Conclusion}
We proposed an algorithm for estimating the right and left subspaces for large MIMO systems, exploiting echoing and the inherent reciprocity in TDD MIMO channels. We first detailed the algorithm within the context of conventional MIMO systems, and then extended it to fit the many operational constraints of the hybrid architecture. Moreover, we highlighted the importance of the subspace decomposition problem, and provided an iterative algorithm for that purpose. Finally, our simulations showed that the high-SNR  performance of our proposed approach it quite similar to their fully digital counterparts.

\addcontentsline{toc}{chapter}{Bibliography}
\bibliographystyle{ieeetr}
\bibliography{ref.bib}

\begin{thebibliography}{1}

\bibitem{Sayeed_CAP_10}
A.~Sayeed and N.~Behdad, ``Continuous aperture phased {MIMO}: Basic theory and
  applications,'' in {\em Communication, Control, and Computing (Allerton),
  2010 48th Annual Allerton Conference on}, pp.~1196--1203, Sept 2010.

\bibitem{Venkateswaran_analogBF_10}
V.~Venkateswaran and A.-J. van~der Veen, ``Analog beamforming in {MIMO}
  communications with phase shift networks and online channel estimation,''
  {\em Signal Processing, IEEE Transactions on}, vol.~58, pp.~4131--4143, Aug
  2010.

\bibitem{Ayach_Spatially_14}
O.~El~Ayach, S.~Rajagopal, S.~Abu-Surra, Z.~Pi, and R.~Heath, ``Spatially
  sparse precoding in millimeter wave {MIMO} systems,'' {\em Wireless
  Communications, IEEE Transactions on}, vol.~13, pp.~1499--1513, March 2014.

\bibitem{Alkhateeb_channel_2014}
A.~Alkhateeb, O.~El~Ayach, G.~Leus, and R.~Heath, ``Channel estimation and
  hybrid precoding for millimeter wave cellular systems,'' {\em Selected Topics
  in Signal Processing, IEEE Journal of}, vol.~8, pp.~831--846, Oct 2014.

\bibitem{Ghauch_BSE_journ}
H.~Ghauch, T.~Kim, M.~Bengtsson, and M.~Skoglund, ``Subspace estimation and
  decomposition for hybrid analog-digital millimetre-wave {MIMO} systems,''
  {\em Manuscript in preparation, preprint available at
  https://www.kth.se/profile/ghauch/}.

\bibitem{Dahl_blind_04}
T.~Dahl, N.~Christophersen, and D.~Gesbert, ``Blind {MIMO} eigenmode
  transmission based on the algebraic power method,'' {\em Signal Processing,
  IEEE Transactions on}, vol.~52, pp.~2424--2431, Sept 2004.

\bibitem{Saad_Numerical_11}
Y.~Saad, ``{Numerical Methods for Large Eigenvalue Problems},'' {\em Manchester
  University Press}, no.~Second Edition, pp.~1--337, 2011.

\bibitem{Baum_sounding_11}
D.~Baum and H.~Bolcskei, ``Information-theoretic analysis of {MIMO} channel
  sounding,'' {\em Information Theory, IEEE Transactions on}, vol.~57,
  pp.~7555--7577, Nov 2011.

\bibitem{3GPP_tr25996}
``Spatial channel model for multiple input multiple output ({MIMO})
  simulations,'' {\em 3GPP TR 25.996 V10.0}, Mar 2011.

\end{thebibliography}

\end{document}